# Plasmon-driven motion of an individual molecule

Tzu-Chao Hung, Brian Kiraly, Julian H. Strik, Alexander A. Khajetoorians, and Daniel Wegner*

Institute for Molecules and Materials, Radboud University, Nijmegen, The Netherlands

*corresponding Author: d.wegner@science.ru.nl

**We demonstrate that nanocavity plasmons generated a few nanometers away from a molecule can induce molecular motion. For this, we study the well-known rapid shuttling motion of zinc phthalocyanine molecules adsorbed on ultrathin NaCl films by combining scanning tunneling microscopy (STM) and spectroscopy (STS) with STM-induced light emission. Comparing spatially resolved single-molecule luminescence spectra from molecules anchored to a step edge with isolated molecules adsorbed on the free surface, we found that the azimuthal modulation of the Lamb shift is diminished in case of the latter. This is evidence that the rapid shuttling motion is remotely induced by plasmon-exciton coupling. Plasmon-induced molecular motion may open an interesting playground to bridge the nanoscopic and mesoscopic worlds by combining molecular machines with nanoplasmonics to control directed motion of single molecules without the need for local probes.**

Controlled activation of molecular motion and conformational changes is vital for creating molecular devices, motors and machines[1,2]. To this end, molecular motion can be induced by a variety of physical and chemical stimuli, depending on the desired functionality[3]. While non-local stimuli, such as an external light source, can address many molecular machines simultaneously[4-6], it is ultimately desirable to create molecular devices where molecules can be addressed individually[7-11]. Controlled on-surface manipulation of single atoms and molecules has been successfully demonstrated using the scanning tunneling microscope (STM), typically via electronic or inelastic (vibrational or vibronic) excitation mechanisms[8-10,12-18]. Within the tip-sample junction, nanocavity plasmons can also be excited[19], as seen in STM-induced light emission (STM-LE)[20-23]. However, to date there has been no demonstration that local nanocavity plasmon-molecule coupling can also be utilized to induce single-molecule motion.

Phthalocyanine (Pc) molecules and their metal complexes on ultrathin NaCl films serve as a model platform to fundamentally investigate optoelectronic phenomena and local interactions on the single molecule scale using STM-LE, including recent discoveries of single-photon emission[24], up-conversion electroluminescence[25], resonance energy transfer[26], coherent dipole coupling and superradiance[27,28], coherent plasmon-exciton interactions[29,30], transient charging and electrofluorochromism[31,32], vibronic spectroscopy[29,31,33,34], tautomerization tracking[34] and doublet emission[35]. Zinc phthalocyanine (ZnPc) is particularly interesting to investigate the possibility of plasmon-induced molecular motion. On the one hand, a rapid shuttling motion between two equivalent adsorption configurations can be induced[26-28,31,33,36-42]. On the other hand, nanocavity plasmons in the tip-sample junction can remotely couple to the molecule, creating an exciton and eventually leading to light emission[26,29-31,34,43,44]. This coherent coupling leads to a Lamb shift of the optical transition, which was found to vary with azimuthal angle of the STM tip along a circumferential path around the molecule, reflecting the ZnPc symmetry[30]. However, thus far there has been no indication that this remote plasmon-exciton coupling may also induce the rapid shuttling motion.

Here, we show that plasmons generated in a nanocavity can induce molecular motion. We quantify this effect, using a combination of STM, scanning tunneling spectroscopy (STS) and STM-LE. We compare two types of electronically equivalent ZnPc molecules, as seen by STS, one being isolated species which can exhibit rapid shuttling and the other being molecules that are anchored to NaCl step edges in order to suppress the shuttling. By comparing the azimuthal dependence of STM-LE spectra of these two ZnPc types, we observe a remarkable difference: the azimuthal modulation of the Lamb shift for anchored molecules is much larger than for isolated ones. Our findings can be understood in a straight-forward model, revealing that rapid shuttling of isolated molecules is not only induced when directly tunneling into the ZnPc but even at remote distances a few nanometers away, where only plasmons couple to the molecule. The fact that nanocavity plasmons can induce single-molecule motion opens up an unexplored route toward inducing and controlling molecular motion by using prepatterned nanoplasmonic structures.

Figure 1(b) shows a constant-current STM image of a typical isolated ZnPc adsorbed on a free terrace of 3 ML NaCl/Ag(111) (see schematic in Fig. 1(a)) when resonantly tunneling out of the highest occupied molecular orbital (HOMO). The HOMO hosts eight lobes (two at each of the four ligand phenyl groups) (see Supplementary Information Section S2)[28]. Yet, in the STM image 16 lobes can be observed, due to a superposition of two meta-stable orientations on the same adsorption site. This apparent superposition results from the rapid shuttling of the molecule about the Zn atom between two metastable orientations, where the phenyl groups are oriented either +11° or −11° with respect to the NaCl ⟨110⟩ crystallographic directions[36] (see Fig. 1(a)). It is known that rapid shuttling occurs when electrons resonantly tunnel either out of the HOMO or into the lowest unoccupied molecular orbital (LUMO)[24,27,28,30,31,33,36]. For voltages within the HOMO-LUMO gap, the molecule can be stabilized in one of the two aforementioned orientations[36,38,39].

In order to quantify the role of the nanocavity plasmons on this rapid shuttling, we first create a reference molecule that does not exhibit shuttling motion. While defects in the NaCl substrate can pin molecules and suppress the shuttling motion[36], we found that ZnPc pinned to defects can exhibit strongly modified electronic and/or optical properties (see Supplementary Information Section S3). Therefore, we used STM manipulation[28] to move an isolated molecule to the edge of a NaCl step, between 3 and 4 ML. We only used binding sites free of local adsorbates or apparent defects, oriented along the ⟨100⟩ direction. Fig. 1(c) presents the constant-current STM image of the same ZnPc shown in (b) after the aforementioned manipulation procedure. The adsorption site is identical to that of isolated molecules, with the Zn atom located above a Cl atom. However, at the imaging voltage in which the isolated molecule exhibits shuttling motion, only eight lobes, correlated with the HOMO, were observed for molecules relocated to the step edge. In comparison with the isolated molecule, the relocated molecule was stabilized such that the phenyl groups align ca. 2-3 degrees off the NaCl ⟨110⟩ directions, as shown in Fig. 1(a). In the following, we refer to this molecule as anchored ZnPc.

Before comparing the optical properties of isolated and anchored ZnPc, we first demonstrate that the local density of states (LDOS) is identical in both cases. In Fig. 1(d), STS spectra are taken for both molecules at locations marked in (b,c). In both cases, the HOMO (i.e., the positive ion resonance[45]) is found at –2.330 eV below the Fermi energy ($E_F$), and the LUMO (negative ion resonance) is located at 1.050 eV above $E_F$. Likewise, the spatial maps of the HOMO and LUMO do not illustrate any modifications in the spatial LDOS distribution compared to the isolated molecule apart from the blurring due to the rapid shuttling of the latter (see Supplementary Information Section S2). We note that it is important to ensure a clean surface and step edge, as even defects a few nanometers away from the molecule can alter the measured d$I$/d$V_S$ of the molecule (see Supplementary Information Section S3). Previous work indicates that a different adsorption site would be accompanied by a different STS spectrum[35], which is further evidence that the molecular adsorption site in both cases is a Cl site. Hence, we conclude that the relocated molecule is not chemisorbed to the step edge, but weakly physisorbed. We speculate that the anchoring of the molecule most likely occurs due to additional van der Waals interactions, which suppress the shuttling motion without altering the internal structural and electronic properties of the molecule. Therefore, in the subsequent discussion, we treat the isolated and anchored molecule as electronically and chemically equivalent.

We now compare the STM current-induced fluorescence of isolated vs. anchored ZnPc, first focusing on the overall spectral features. As previously reported, the STM-LE spectrum on top of the isolated ZnPc (blue curve in Fig. 1(e)) illustrates two features: a broad peak at 1.902 eV, and a shoulder at 1.911 eV[24,27,28,30,31,33,36]. While the former has been interpreted as the Q(0,0) singlet transition fluorescence peak, it has been suggested that the latter is related to the shuttling motion of the molecule, as it is not observed within aggregated molecule chains[27,28,33]. In comparison, the STM-LE spectrum on top of the anchored molecule (red curve in Fig. 1(e)) shows a single peak located at a different energy (1.910 eV) with a much narrower width and no shoulder feature. We attribute this peak to the Q(0,0) transition. Its blueshift with respect to isolated molecule can be explained by the different dielectric environment defined by the step edge and/or by additional van der Waals interactions, leading to a modified exciton binding energy[28]. Interestingly, the asymmetric spectral shape reflects a Fano lineshape, indicating the presence of two coherent excitation channels. The two potential excitation channels discussed in the literature are: (*i*) excitation resulting from direct tunneling into the molecule, i.e. electron and hole injection into frontier orbitals, leading to the formation of an exciton[28,36]; (*ii*) inelastic excitation resulting from tunneling between tip and substrate, exciting nanocavity plasmons, and permitting a plasmon-induced energy transfer and exciton formation in the molecule[31,33]. While previous studies concluded that either (*i*) or (*ii*) alone is responsible for the observed STM-LE spectrum for the isolated molecule, our finding clearly shows that both excitation mechanisms are activated and couple coherently, resulting in the observed Fano resonance[30,43,44]. This conclusion can be made due to the sharper spectral feature and clear lineshape presented for the anchored molecule. Therefore, it is important to elucidate a possible impact of nanocavity plasmons on inducing the shuttling motion observed for isolated molecules.

In order to quantify the role of nanocavity plasmons in molecule-plasmon coupling, we focus on the Lamb shift. The Lamb shift is a redshift in the molecular optical response due to a coherent coupling of the discrete molecular emitter state with the continuum of states of the nanocavity plasmons created in the tip-sample junction[30,46]. We employed it as a highly sensitive remote probe for the molecular orientation of the ZnPc with respect to the NaCl substrate. This is possible because the Lamb shift was recently found to be modulated along a circle around the molecule with a π/2 periodicity in the azimuthal direction. This Lamb shift modulation (LSM) was attributed to the $D_{4h}$ symmetry of the ZnPc with its two orthogonal transition dipole moments along the main molecular axes, leading to an anisotropic coupling of the molecule with the nanocavity plasmons. However, a rapid shuttling would diminish the detected effective LSM, as a superposition of two phase-shifted LSMs from the two metastable orientations of the molecule would be observed. So far, STM-LE based Lamb shift measurements were only performed remotely, with the tip a few nanometers away from the molecule[30], but no azimuthal modulation of the Lamb shift has been reported when directly tunneling into the isolated molecule, presumably because it may be invisible due to the rapid shuttling and the ensuing spectral broadening.

To acquire a reference of the LSM for ZnPc both on and off the molecule, we used the STM-LE spectra from an anchored ZnPc. Fig. 2(a) presents a series of false-color plotted STM-LE spectra of the anchored ZnPc at a radial distance from the molecular center (i.e. the Zn atom) of $r$ = 825 pm (i.e. on the Pc ligand) at different azimuthal angles $\theta$. Here, $\theta$ = 0° is defined along the crystallographic NaCl $[0\bar{1}0]$ direction as shown in Fig. 1(a). To maintain comparable nanocavity plasmons for each $\theta$, the STM-LE spectra were recorded in constant-height mode. The resulting series of STM-LE spectra revealed that the emission peak position approximately oscillates with a π/2 periodicity owing to the azimuthal modulation of the Lamb shift. The nonequal energy level shift at 45° and 135° can be explained by an asymmetric tip shape (see Supplementary Information Section S4 for more details). In Fig. 2(b), STM-LE spectra at $\theta$ = 0° (blue curve) and at $\theta$ = 45° (orange) are shown. The Lamb shift is largest at $\theta$ = 45°, leading to a stronger redshift with increased molecule-plasmon coupling compared to the STM-LE spectrum at $\theta$ = 0°. A fit of the STM-LE spectra with a Fano function (see Supplementary Information Section S5) reveals comparable widths and $q$ factors, with a LSM of $\Delta E_{\text{LS}} = E_0(0°) - E_0(45°) =$ 3.9 ± 0.3 meV. In comparison, Zhang et al. reported a much smaller LSM of $\Delta E_{\text{LS}} \approx 2$ meV for isolated molecules[30]. The latter was not measured on but next to the molecule ($r$ ≥ 1.6 nm), and the Lamb shift is known to exhibit a lateral tip-molecule distance dependence[30], which may explain the larger LSM measured here at $r$ = 825 pm. To test this, we also performed experiments at various distances. Fig. 2(c) shows two STM-LE spectra obtained at $r$ = 1810 pm (i.e., the tip was positioned next to the molecule) at an angle of 0° (blue) and 45° (orange), respectively. At this distance, direct tunneling into the molecule is not possible (hence excitation channel (*i*) can be completely ruled out) and the molecular resonance is only observed indirectly via plasmon-exciton coupling, leading to a faint Fano dip in the plasmon resonance STM-LE spectrum[30,43,44]. Irrespective of the small signal-to-noise ratio, these remote STM-LE spectra also display a clear LSM, which we quantified as $\Delta E_{LS}$ = 5.2 ± 0.5 meV. Additional STM-LE series at various radial distances confirm this observation (cf. inset of Fig. 4(a), and Supplementary information Fig. S8).

In order to rule out that the enhanced LSM of the anchored molecule originates from any other artifacts, we performed the same angular-dependent measurements next to the isolated molecule on 3 ML NaCl/Ag(111) with the identical tip used in Fig. 2 as well as the same stabilization parameters. Fig. 3(a) shows two STM-LE spectra for a radial distance of $r$ = 1835 pm at $\theta$ = 0° and 45°, respectively (where $\theta$ is defined as before, see Fig. 1(a)). Again, a π/2 periodicity can be observed as was seen in Fig. 2. Likewise, the unequal energy shift of 45° and 135° presumably originated from the same asymmetric tip. Fitting the 0° and 45° STM-LE spectra with Fano profiles, we found $\Delta E_{LS}$ = 2.5 ± 0.3 meV. This is consistent with the value previously reported[30], and confirms that the LSM of isolated ZnPc is only about half of the value seen for the anchored molecule. We also tested whether the LSM can be observed for isolated molecules when the tip is positioned above the molecule. Fig. 3(c) shows two STM-LE spectra at a radial distance of $r$ = 825 pm, which reproduces the STM-LE spectrum of isolated ZnPc in Fig. 1(e). However, due to the spectral broadening of the Q(0,0) resonance peak and the additional shoulder feature, no clear difference between the STM-LE spectra could be observed.

For a statistically better quantification of the LSM, we determined the Q(0,0) resonance energies at various $\theta$ and $r$. Fig. 4(a) summarizes data for the anchored molecule, both above the ZnPc ligand ($r$ = 825 pm, orange triangles) and for the remote measurement ($r$ = 1810 pm, blue circles). The LSM remains comparably large. All datasets can be phenomenologically described as a periodic pattern using the function[30] $E_0(\theta) = A \cos 4\theta + B$, where a statistical value of the LSM is given by $2A = \Delta E_{\mathrm{LS}}$ and the offset $B$ is the angle-averaged resonance energy. Fitting the data, we found 2$A$($r$ = 825 pm) = 3.6 ± 0.4 meV and 2$A$($r$ = 1810 pm) = 4.8 ± 0.6 meV. The radial distance dependence (inset of Fig. 4(a)) shows that $\Delta E_{\mathrm{LS}}$ decreases when the tip approaches the ZnPc center, reflecting the lateral dimensions of the nanoplasmonic cavity and the molecule; for symmetry reasons, the LSM must vanish for $r$ = 0[30].

As the LSM measurements of the anchored and the isolated ZnPc were performed under identical conditions (same tip and similar vertical and lateral tip-molecule distances used in Fig. 2(c), 3(b)), a quantitative comparison is possible. Fig. 4(b) summarizes the angle-dependent remote measurements for the isolated molecule. The significantly diminished LSM can neither be attributed to a reduced coupling of the nanocavity plasmons to the ZnPc molecule, nor to any other experimental artifact. This implies that remote measurements of isolated molecules are more invasive than previously suggested and that the plasmon-exciton coupling alone may induce the shuttling motion of the isolated ZnPc molecule. We can model an expected LSM of isolated ZnPc that is rapidly shuttling during a measurement by a superposition $E_0(\theta) = \frac{1}{2}(E_0^+ + E_0^-) + B$ of the LSMs originating from the two orientation angles, $E_0^\pm(\theta) = A\cos 4(\theta \pm 11°)$. We used an amplitude $2A$ = 4.8 meV that corresponds to the LSM of anchored ZnPc at $r$ = 1810 pm. The resulting model curve is displayed as solid black line in Fig. 4(b). Despite its simplicity, the model describes the observed LSM of isolated ZnPc quite well. Hence, we conclude that the nanocavity plasmons generated a few nanometers away from ZnPc can induce the rapid shuttling motion.

A recent study on various metal phthalocyanines concluded that the rapid shuttling motion upon direct tunneling into the molecule involves an intermediate charged molecular state[39,42]. Our results provide evidence that in the case of remote measurements, i.e. in the absence of direct transport through the molecule, plasmon-exciton coupling induces the shuttling. It remains unclear how exactly this excitation couples to the mechanical degrees of freedom. A recent theoretical treatment of a related molecule (MgPc) found an energy barrier between the two adsorption orientations of 9 meV[38], which interestingly corresponds to the energy difference observed between the Q(0,0) transition and the shoulder in STM-LE of isolated ZnPc (Fig. 1(e) and Fig. 3(c)). Such a barrier can easily be overcome if the exciton couples to a vibrational or hindered rotational mode.

In summary, we studied the influence of the rapid shuttling motion on the fluorescence properties of ZnPc on the NaCl surface, by comparing the LSMs of an anchored and an isolated molecule. For the latter, we found a diminished LSM, which provides evidence that isolated molecules rapidly shuttle during remote measurements due to plasmon-molecule coupling. Our results should provide valuable input for future theoretical treatments to identify the exact mechanism that leads to molecular motion based on coupling to nanocavity plasmons. A better understanding could also offer a route toward inducing lateral or even directed motion of accordingly designed molecules on surfaces. Even more, it may provide a novel playground for controlling the motion of molecular machines such as rotors, motors or nanocars[7,9,14,47], without the need for a local probe by utilization of nanoplasmonic structures[48], which can readily be made using state-of-the-art lithography techniques – an intriguing prospect toward scalable control of complex molecular machines.

**Methods**

We performed the STM and STM-LE measurements using an Omicron ultra-high vacuum low-temperature STM system operated at $T$ = 4.5 K with a base pressure below $1 \cdot 10^{-10}$ mbar[49]. All experiments shown here were performed for ZnPc molecules adsorbed on 3 ML NaCl/Ag(111) (for additional data on 2 ML NaCl/Ag(111), see Supplementary Information Section 4). All STM-LE spectra presented in this paper were acquired using the same plasmonic tip, shown in Supplementary Information Section 1, and measured on the same ZnPc molecule, in order to eliminate tip-dependent variations in the STM-LE spectra. All relative changes reported for a given STM-LE spectrum were reproducible with different microtips (see, e.g., Supplementary Information Section 4).

An adapted photon collection apparatus was used in our experiment[50]. An *in-situ* lens (diameter = 10.0 mm, focal length = 15.0 mm, VIS-NIR coated, plano-convex, Edmund Optics Ltd.) was placed ca. 15 mm away from the tunnel junction with an angle of 25° to the surface to collimate the photons and subsequently guide them through the cryoshield and UHV viewports. An *ex-situ* lens (diameter = 25.0 mm, focal length = 50.0 mm, uncoated, Thorlabs) redirected the light into an optical fiber bundle which was connected to a grid spectrometer (SpectraPro HRS-300, Princeton Instruments) equipped with a $LN_2$-cooled CCD detector (PyLoN 100, Princeton Instruments). The entrance slit to the spectrometer was set to a width of 200 μm for all the reported spectra. Three different gratings were used in our experiment. The low spectral resolution grating (150 grooves/mm, 10.6 meV) was used to characterize the plasmon resonance of the tip. The mid (600 grooves/mm, 2.5 meV) and high (1200 grooves/mm, 1.3 meV) spectral resolution gratings were used when characterizing the main transition peak of the molecules.

For the sample preparation, a Ag(111) single crystal (MaTeck) was cleaned by multiple cycles of sputtering and annealing, followed by deposition of NaCl from a Knudsen cell while the Ag(111) surface was kept at room temperature. The sample was then transferred into the cryogenic STM and cooled to base temperature. ZnPc molecules (97%, Sigma-Aldrich) were thermally evaporated from a Knudsen cell and deposited onto the cold NaCl/Ag(111) sample (T < 6 K) inside the STM. STM images were taken both in constant-current and constant-height mode (as stated), with d$I$/d$V_S$ spectra as well as d$I$/d$V_S$ maps obtained using a lock-in technique with a modulation voltage $V_{rms}$ = 5 mV at a frequency of 819 Hz.

To enhance the optical STM-LE output, we created Ag-coated W tips by controllably indenting flashed W tips into the Ag(111) crystal[33], followed by further modification through dipping and voltage pulsing on bare Ag to achieve the desired plasmon resonance (i.e. around 1.9 eV, close to the energy of the Q(0,0) transition). The photon collection efficiency was determined to be sufficient when the maximum intensity of the plasmon resonance spectrum acquired on 3 ML NaCl/Ag(111) reached more than 100 counts using feedback parameters of $V_S$ = -2.5 V, $I$ = 100 pA and an acquisition time of $t$ = 60 s). Supplementary Fig. S1 shows a typical plasmon resonance spectrum after such optimization. This particular spectrum corresponds to that of the tip used for all STM-LE spectra shown in the main text.

33  Doppagne, B. *et al.*, Vibronic Spectroscopy with Submolecular Resolution from STM-Induced Electroluminescence, *Phys Rev Lett* **118** (2017).
34  Doppagne, B. *et al.*, Single-molecule tautomerization tracking through space- and time-resolved fluorescence spectroscopy, *Nat Nanotechnol* **15**, 207 (2020).
35  Doležal, J. *et al.*, Mechano-Optical Switching of a Single Molecule with Doublet Emission, *Acs Nano* **14**, 8931 (2020).
36  Dolezal, J. *et al.*, Charge Carrier Injection Electroluminescence with CO-Functionalized Tips on Single Molecular Emitters, *Nano Lett* **19**, 8605 (2019).
37  Kaiser, K., Gross, L. & Schulz, F., A Single-Molecule Chemical Reaction Studied by High-Resolution Atomic Force Microscopy and Scanning Tunneling Microscopy Induced Light Emission, *Acs Nano* **13**, 6947 (2019).
38  Miwa, K., Imada, H., Kawahara, S. & Kim, Y., Effects of molecule-insulator interaction on geometric property of a single phthalocyanine molecule adsorbed on an ultrathin NaCl film, *Phys Rev B* **93** (2016).
39  Patera, L. L., Queck, F., Scheuerer, P., Moll, N. & Repp, J., Accessing a Charged Intermediate State Involved in the Excitation of Single Molecules, *Phys Rev Lett* **123** (2019).
40  Schaffert, J. *et al.*, Imaging the dynamics of individually adsorbed molecules, *Nat Mater* **12**, 223 (2013).
41  Tao, M. L. *et al.*, Structural transitions in different monolayers of cobalt phthalocyanine film grown on Bi(111), *J Phys D Appl Phys* **49** (2016).
42  Peller, D. *et al.*, Sub-cycle atomic-scale forces coherently control a single-molecule switch, *Nature* **585**, 58 (2020).
43  Kroger, J., Doppagne, B., Scheurer, F. & Schull, G., Fano Description of Single-Hydrocarbon Fluorescence Excited by a Scanning Tunneling Microscope, *Nano Lett* **18**, 3407 (2018).
44  Nian, L. L., Wang, Y. F. & Lu, J. T., On the Fano Line Shape of Single Induced by a Scanning Tunneling Molecule Electroluminescence Microscope, *Nano Lett* **18**, 6826 (2018).
45  Repp, J., Meyer, G., Stojkovic, S. M., Gourdon, A. & Joachim, C., Molecules on insulating films: Scanning-tunneling microscopy imaging of individual molecular orbitals, *Phys Rev Lett* **94** (2005).
46  Lamb, W. E. & Retherford, R. C., Fine Structure of the Hydrogen Atom by a Microwave Method, *Phys Rev* **72**, 241 (1947).
47  Kassem, S. *et al.*, Artificial molecular motors, *Chem Soc Rev* **46**, 2592 (2017).
48  Fang, Y. R. & Sun, M. T., Nanoplasmonic waveguides: towards applications in integrated nanophotonic circuits, *Light-Sci Appl* **4** (2015).
49  Kiraly, B. *et al.*, An orbitally derived single-atom magnetic memory, *Nat Commun* **9**, 3904 (2018).
50  Keizer, J. G., Garleff, J. K. & Koenraad, P. M., Simple and efficient scanning tunneling luminescence detection at low-temperature, *Rev Sci Instrum* **80** (2009).
16

**ACKNOWLEDGMENT**

This work is part of the research program of the Foundation for Fundamental Research on Matter (FOM), which is financially supported by the Netherlands Organization for Scientific Research (NWO).

**Author Contributions**

T.C.H. and D.W. designed and T.C.H., B.K. and J.H.S. conducted the experiments. T.C.H., J.H.S., B.K. and D.W. analyzed the data. All authors contributed to discussions and writing of the manuscript and have given approval to the final version of the manuscript.

**Competing Interests**

The authors declare no competing interests.

**Materials and Correspondence**

Correspondence to Daniel Wegner.

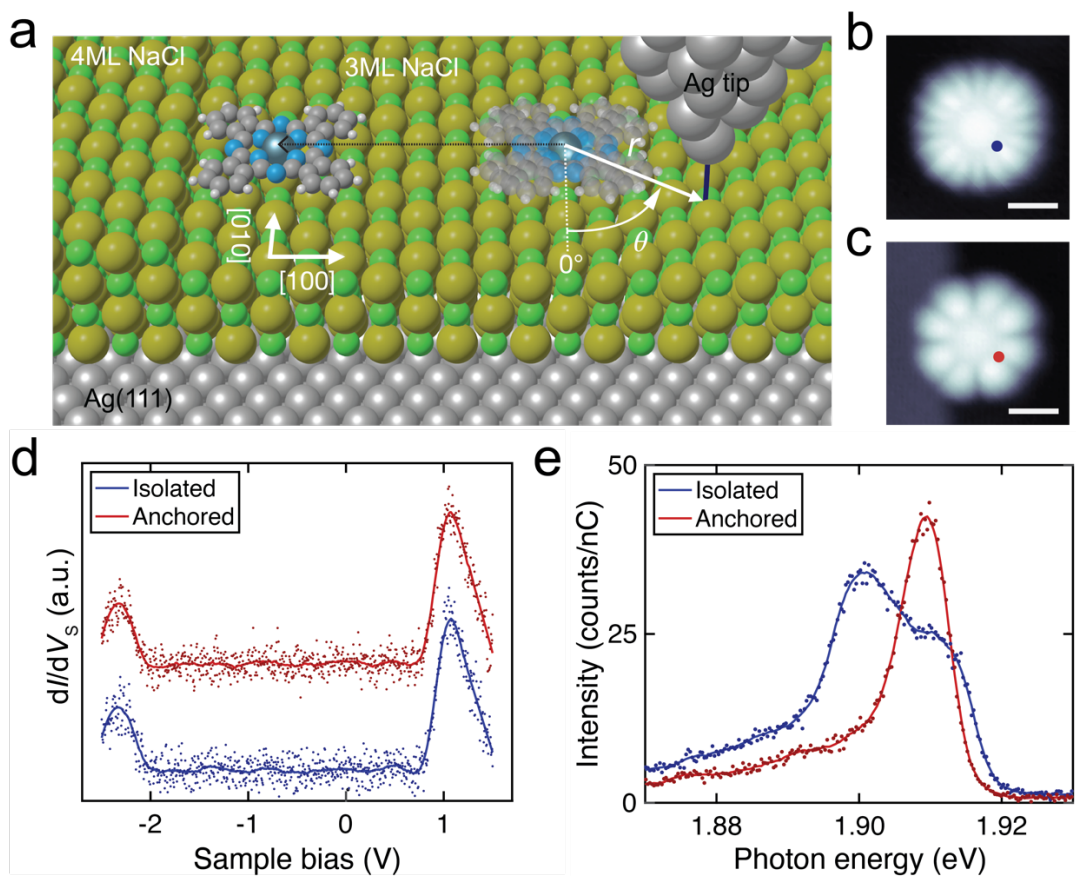

**Figure 1 | Isolated *vs.* anchored ZnPc on 3 ML NaCl/Ag(111). a,** Schematics of STM-LE and manipulation experiments. Structural model illustrates the ±11° shuttling between two adsorption orientations for isolated ZnPc (right), which is suppressed for anchored ZnPc (left). The lateral molecule-tip distance $r$ and azimuthal angle $\theta$ are defined from the ZnPc center, with $\theta = 0°$ defined downward along the NaCl $[0\bar{1}0]$ direction as displayed. **b,** Constant-current STM image of an isolated ZnPc before manipulation, and **c,** anchored ZnPc after moving it to a 4 ML NaCl step edge ($V_S = –2.5$ V, $I = 5.5$ pA, scale bars = 1 nm). **d,** STS of the isolated and anchored ZnPc. **e,** STM-LE spectra of the isolated and the anchored ZnPc ($V_S = -2.5$ V, $I = 100$ pA, acquisition time $t = 180$ s) measured at the positions marked in (**b**) and (**c**). Solid lines in (**d**) and (**e**) are 80-point and 20-point Savitzky-Golay filtered curves of the raw data respectively.

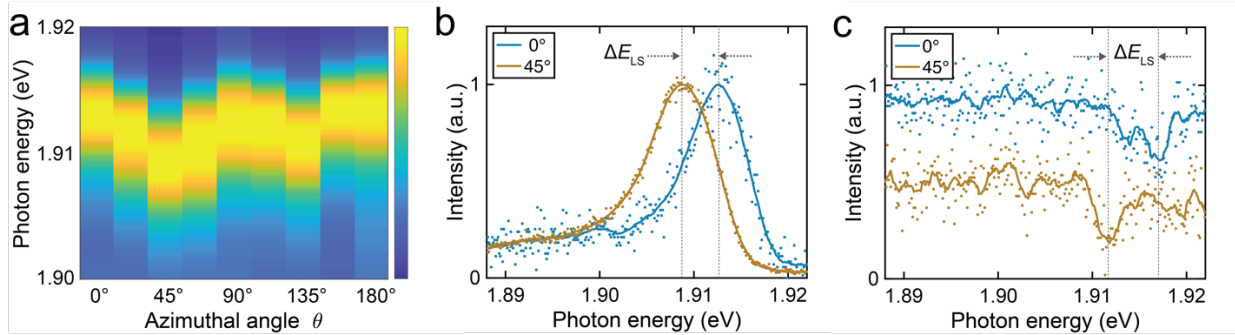

**Figure 2 | Lamb shift modulation (LSM) of anchored ZnPc. a,** Color plot of normalized STM-LE spectra showing periodic modulation of the Lamb shift. **b,** Two typical Fano-peak STM-LE spectra taken on top of anchored ZnPc ($r$ = 825 pm) for $\theta$ = 0° and 45°, with $\Delta E_{\text{LS}}$ = 3.9 ± 0.3 meV (taken in constant-height mode with feedback loop opened at the ZnPc center, $V_S$ = –2.5 V, $I$ = 500 pA, $t$ = 120 s). **c,** Two typical Fano-dip STM-LE spectra taken next to anchored ZnPc ($r$ = 1810 pm) at $\theta$ = 0° and 45°, with $\Delta E_{\text{LS}}$ = 5.2 ± 0.5 meV (constant-current mode $V_S$ = -2.5 V, $I$ = 200 pA, $t$ = 60 s). Solid lines in (**b**) and (**c**) are 50-point and 20-point Savitzky-Golay filtered curves of the raw data respectively.

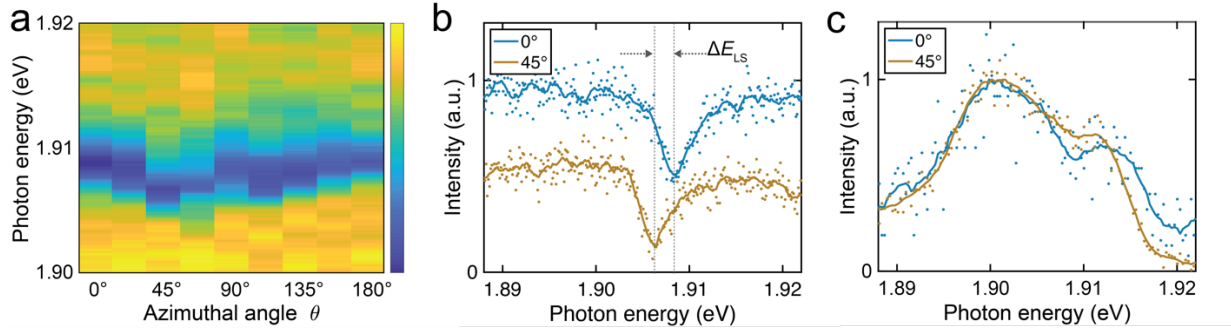

**Figure 3 | LSM of isolated ZnPc. a,** Color plot of normalized STM-LE spectra showing periodic modulation of the Lamb shift. **b,** Two typical Fano-dip STM-LE spectra taken next to isolated ZnPc ($r$ = 1835 pm) at $\theta$ = 0° and 45°, with $\Delta E_{LS}$ = 2.5 ± 0.3 meV (constant-current mode, $V_S$ = -2.5 V, $I$ = 200 pA, $t$ = 120 s). **c,** Two STM-LE spectra obtained on top of isolated ZnPc ($r$ = 825 pm) at $\theta$ = 0° and 45°, with no assignable LSM (constant-height mode, feedback opened at ZnPc center, $V_S$ = –2.5 V, $I$ = 100 pA, $t$ = 60 s). Solid lines in (**b**) and (**c**) are 200-point and 20-point Savitzky-Golay filtered curves of the raw data respectively.

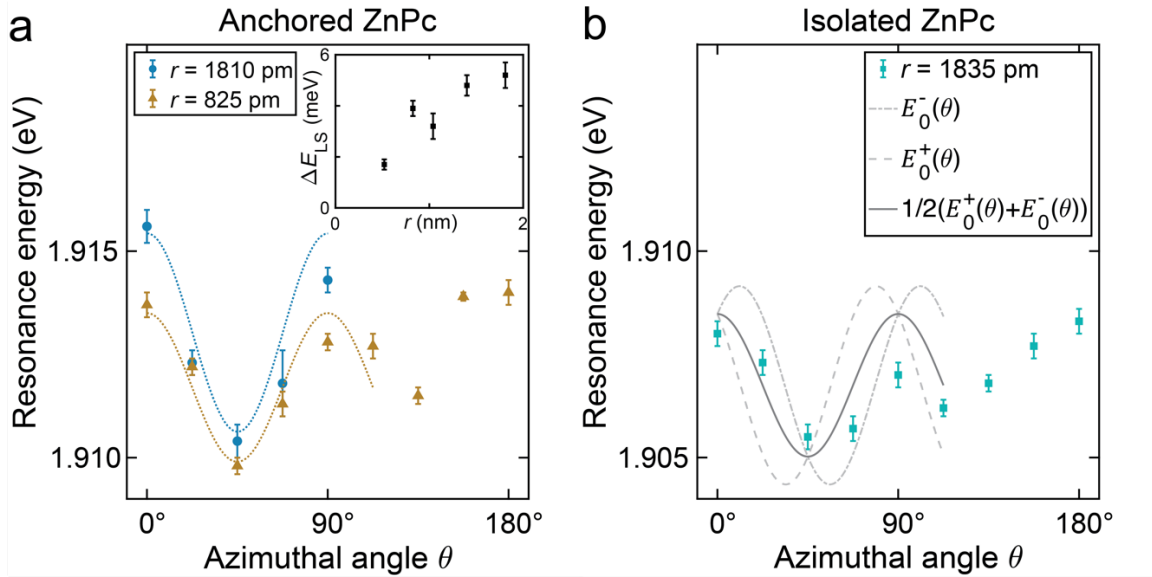

**Figure 4 | Analysis of the LSM for anchored *vs.* isolated ZnPc. a,** Azimuthal angle dependence of the Q(0,0) resonance for anchored ZnPc at *r* = 825 pm (orange triangles) and *r* = 1810 pm (blue circles). A cosine fit yields peak-to-peak amplitudes $2A = \Delta E_{LS}$. The inset shows a summary of the radius-dependent $\Delta E_{LS}$ for various *r*. **b,** Azimuthal angle dependence of the Q(0,0) resonance for isolated ZnPc at *r* = 1835 pm (light-blue squares). The gray dashed lines are model curves based on results from (**a**) for assumed azimuthal angles of −11° and +11°, respectively. Their superposition (solid gray line) reflects the expected LSM in case of rapid shuttling. The model agrees well with the raw data.

# Supplementary Information:
# Plasmon-driven motion of an individual molecule

*Tzu-Chao Hung, Brian Kiraly, Julian H. Strik, Alexander A. Khajetoorians, and Daniel Wegner**

*Institute for Molecules and Materials, Radboud University, Nijmegen, The Netherlands*

*Correspondence to: d.wegner@science.ru.nl

**The supplementary information includes:**

**S1 Plasmon resonance spectrum of the tip used for the presented STM-LE data**

**S2 Molecular orbitals of isolated ZnPc and anchored ZnPc**

**S3 STM-LE response of ZnPc due to the presence of defects and the NaCl Moiré pattern**

**S4 Lamb shift modulation of anchored ZnPc on 2 ML NaCl**

**S5 Fitting STM-LE spectra by Fano profiles**

**S1 Plasmon resonance spectrum of the tip used for the presented STM-LE data**

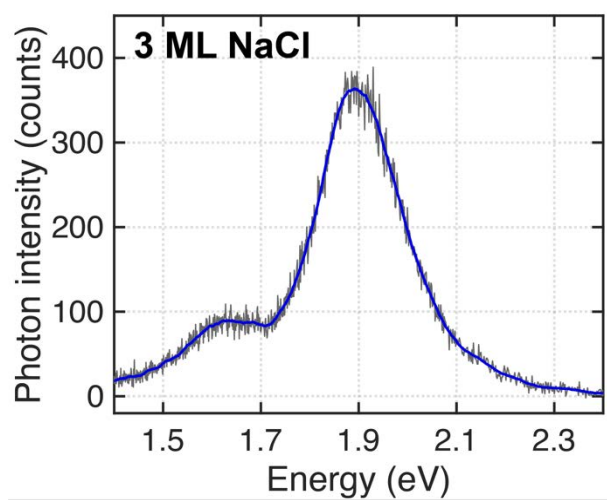

**Figure S1: Plasmon resonance spectrum of the tip used for the presented STM-LE data.** A broad plasmon resonance band, acquired on 3 ML NaCl/Ag(111), was intentionally tuned around 1.9 eV ($V_S$ = -2.5 V, $I_T$ = 100 pA, $t$ = 60 s). Raw data is plotted as gray solid line, the blue solid line is a 60-point Savitzky-Golay filtered curve of the raw data.

## S2 Molecular orbitals of isolated ZnPc and anchored ZnPc

The HOMO and degenerate LUMO/LUMO+1 have 4-fold symmetry in the free-standing ZnPc molecule[1,2]. Experimentally, the local density of states (LDOS) of molecular orbitals (MOs) can be observed using *dI/dV* mapping. Fig. S2 shows the constant-height *dI/dV* maps of the HOMO and LUMO of the isolated and anchored ZnPc molecules. The HOMO of the isolated ZnPc is shown in Fig. S2(a), with the expected four-fold symmetry. Alternatively, on ultrathin NaCl one can also visualize the MOs using constant-current or constant-height STM images, as there is negligible LDOS within the HOMO-LUMO gap when molecules are adsorbed on insulators[3]. In Fig. S2(e), the constant-height STM image provides better resolution of the MOs. The observation of 16 distinct lobes (same feature in Fig. 1(b)) in this current map confirms the ±11° shuttling motion of the molecule with respect to the NaCl film[4]. The degenerate LUMO/LUMO+1 reveals the expected 4-fold symmetry in Fig. S2(b) and Fig. S2(f). To ensure the symmetry of the MOs remained the same upon anchoring the molecule to the NaCl step edge, the same measurements were performed on the anchored molecules. Fig. S2(c) shows the HOMO of the anchored molecule, where eight lobes are observed (compared to the 16 from Fig. S2(a)) because the molecule was no longer shuttling (see also Fig. 1(c)). Fig. S2(d) shows the 4-fold symmetry of the LUMO/LUMO+1, which indicates that the presence of the NaCl step edge did not break the degeneracy of these orbitals. The maps in Fig. S2(c, d, g, h) are consistent with the STS data (Fig. 1(d)), which shows that the electronic state of the molecule was not modified when the molecule was anchored to the NaCl step edge.

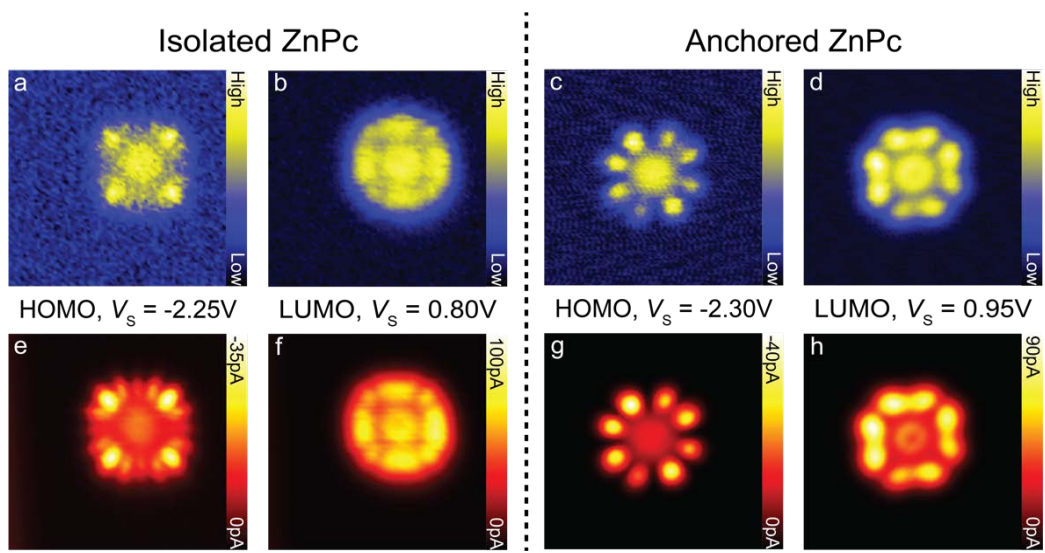

**Figure S2: Molecular orbitals of isolated ZnPc and anchored ZnPc.** (a-d) The constant-height differential conductance maps (4×4 nm$^2$) show (a) the HOMO of the isolated ZnPc ($V_S$ = -2.25V), (b) the LUMO of the isolated ZnPc ($V_S$ = 0.80V), (c) the HOMO of the anchored ZnPc ($V_S$ = -2.30V), and (d) the LUMO of the anchored ZnPc ($V_S$ = 0.95V). (e-h) Simultaneously acquired constant-height STM images corresponding to (a-d). The differential conductance maps were acquired by using the lock-in technique. The bias modulation was $V_{rms}$ = 5 mV at $f_{mod}$ = 437.1 Hz. The feedback loop was opened at the center of the molecule at $V_S$ = -2.5 V and $I_T$ = 100pA for the isolated molecule and $V_S$ = -2.5 V and $I_T$ = 50pA for the anchored molecule.

**S3 STM-LE response of ZnPc due to the presence of defects and the NaCl Moiré pattern**

We found that the optical response of single ZnPc molecules is very sensitive to the surrounding environment. Here, we give two illustrative examples: (1) defects in the vicinity of the anchored the molecule, and (2) the molecule pinned to a point defect. Fig. S3(a) shows a STM constant-current image obtained at negative bias ($V_s$ = -2.3V), with a ZnPc molecule anchored to a step edge. However, when imaging at positive bias ($V_s$ = 2.3V), several defects could be observed in the vicinity of the molecule (Fig. S3(b)). To understand the electronic properties of the molecule, under the influence of these defects, the molecular orbitals (MOs) were mapped using constant-height $dI/dV$ maps. The HOMO of the anchored ZnPc shows 4-fold symmetry in the $dI/dV$ map (Fig. S3(c)). However, in Fig. S3(d), the symmetry of the LUMO/LUMO+1 is reduced from 4-fold to 2-fold, which implies that the degeneracy of the LUMO and LUMO+1 was broken due to the presence of the defect. Furthermore, STM-LE spectra taken at two neighboring phenyl arms of the Pc ligand (tip positions are shown in Fig. S3(a)) reveal that the Q(0,0) transition peak is located at different energies, implying that the two orthogonal transition dipoles of the ZnPc were no longer degenerate, consistent with the 2-fold symmetry seen in the $dI/dV$ map. This example demonstrates that even remote defects (i.e. without direct physical contact) can have an influence on the properties of the ZnPc. We speculate that the defects here are most likely Cl vacancy sites[5]. The breaking of degeneracy of the LUMO/LUMO+1 and the transition dipoles of the ZnPc could be attributed to local electric fields, as the vacancies lead to a local charge accumulation.

The effect of a point defect in the NaCl surface in direct contact with the ZnPc is summarized in Fig. S4. Here, a ZnPc molecule was pinned on top of a defect (presumably also a Cl vacancy) using STM manipulation. The defect-pinned ZnPc showed a severe distortion in the STM constant-current image (Fig. S4(a)), indicating that the electronic properties were strongly influenced by the underlying defect. Moreover, this pinned ZnPc exhibited unusual optical properties (Fig. S4(b)): multiple peaks were observed in the STM-LE spectrum. The spectra were also strongly site-dependent (not shown).

To avoid these complicating issues, we ensured that NaCl step edges were atomically pristine prior to anchoring a ZnPc to it. For the results shown in the main text; the step edges were checked by using constant-current imaging at both polarities. A region was considered clean when no defect was found within a radius of 3 nm around the position where a ZnPc would later be anchored. We note that we also carefully checked the potential impact of the NaCl Moiré pattern on Ag(111) (Fig. S5) on the STM-LE spectra of ZnPc[6], but we essentially found no influence, i.e. STM-LE spectra of isolated ZnPc were always identical (Fig. S5(e)), irrespective of the location of the molecule within the Moiré pattern.

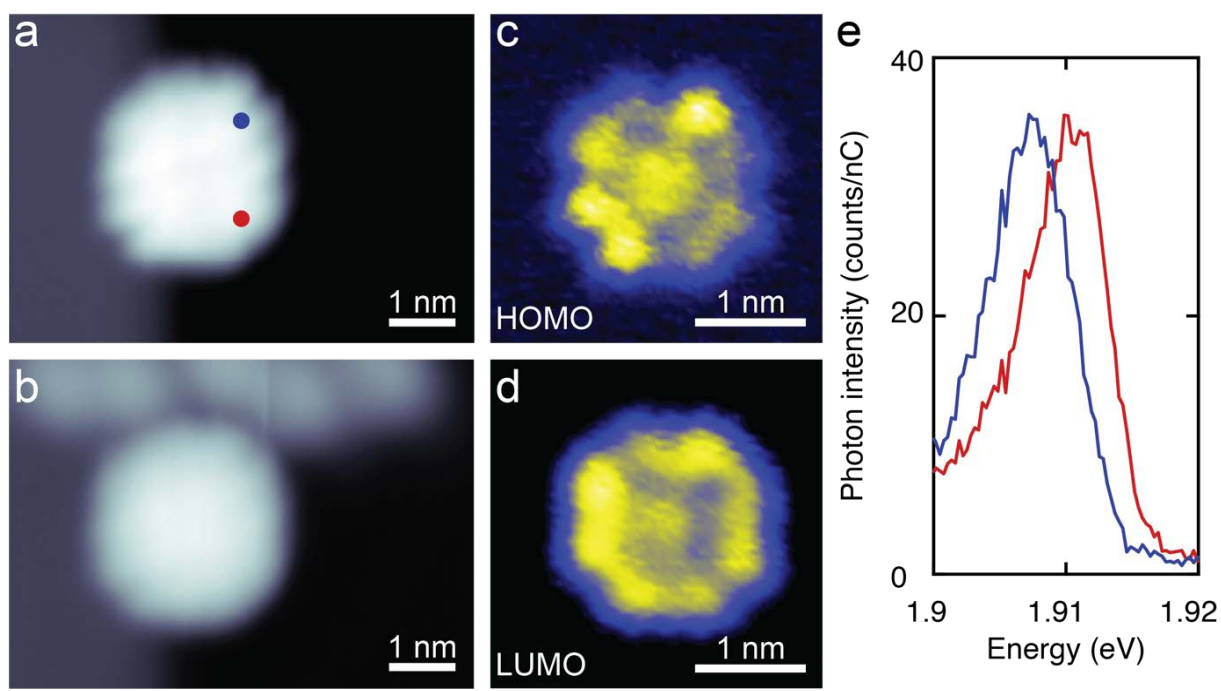

**Figure S3: Degeneracy breaking by defects in the vicinity of an anchored molecule.** (a) Constant-current STM image taken at $V_S$ = -2.3V shows no defect present ($V_S$ = -2.3 V, $I_T$ = 15 pA). (b) Constant-current STM image taken at $V_S$ = 2.3V uncovers defects in the NaCl next to the molecule ($V_S$ = 2.3 V, $I_T$ = 15 pA). (c) Constant-height $dI/dV$ map of the HOMO and (d) the LUMO of the anchored molecule, revealing a two-fold symmetry of the latter. The bias modulation was $V_{rms}$ = 5 mV at $f_{mod}$ = 437.1 Hz. The feedback loop was opened at the center of the molecule at $V_S$ = -2.5 V and $I_T$ = 100pA. (e) STM-LE spectra acquired at the positions marked in (a) show a shift of the fluorescence ($V_S$ = -2.5 V, $I_T$ = 100 pA, $t$ = 180s).

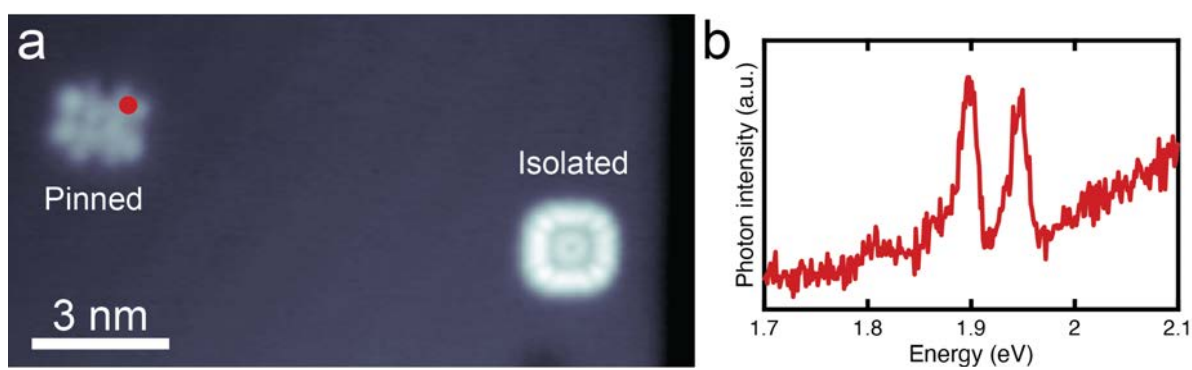

**Figure S4: Defect-pinned ZnPc.** (a) A ZnPc was intentionally manipulated on top of a NaCl surface defect, the constant-current STM image shows the defect-pinned ZnPc and an isolated ZnPc ($V_S$ = 0.8 V, $I_T$ = 15 pA). (b) STM-LE spectrum acquired on the defect-pinned ZnPc (position marked in (a)) revealing a strongly modified fluorescence response ($V_S$ = -2.5 V, $I_T$ = 200 pA, $t$ = 300s).

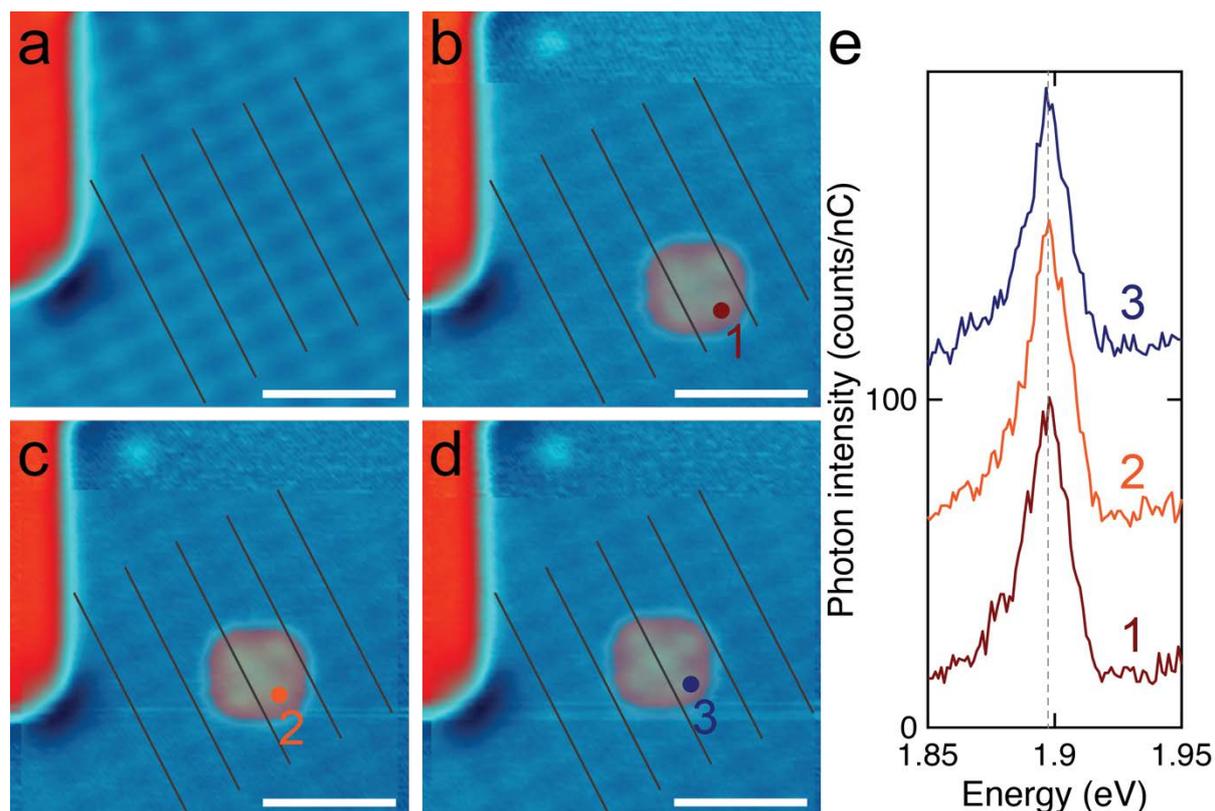

**Figure S5: ZnPc on different adsorption sites of the Morié pattern.** (a) Constant-current STM image of 2 ML NaCl/Ag(111), resolving the structural Morié pattern of the NaCl layer (cf. gray lines to guide the eye) ($V_S$ = 2.5 V, $I_T$ = 20 pA, scale bars = 3 nm). (b-d) A single ZnPc molecule was manipulated onto the area shown in (a) and placed on different locations within the Morié pattern. The images are overlaid onto that shown in (a) with 60% transparency ($V_S$ = -2.5 V, $I_T$ = 6.5 pA). (e) STM-LE spectra of the ZnPc at different adsorption sites, as labelled in (b-d) ($V_S$ = -2.5 V, $I_T$ = 100 pA, $t$ = 30s).

**S4 Lamb shift modulation of anchored ZnPc on 2 ML NaCl**

An orientation-dependent measurement (comparable to Fig. 2) was also done on an anchored ZnPc molecule on 2 ML NaCl/Ag(111) (i.e., anchored to a 3 ML step edge), using a different plasmonic tip. Fig. S6(a) summarizes the STM-LE spectra of the Q(0,0) transition peak as a function of $\theta$ from 0° to 180°, where $r$ = 915 pm. The resulting series of STM-LE spectra reveals that the emission peak position oscillates with a π/2 periodicity. The equal energy level shift at 45° and 135° can be attributed to the tip shape (plasmonic nanocavity) being close to point-symmetric. To analyze this further, we plot STM-LE spectra at $\theta$ = 0° (blue) and at $\theta$ = 45° (orange) (Fig. S5(b)). The Lamb shift is largest at $\theta$ = 45°, which is redshifted relative to the spectrum at $\theta$ = 0°. The spectral Fano profiles and widths are comparable. A fit of the spectra with a Fano function yields a LSM of $\Delta E_{LS}$ = 4.2 ± 0.3 meV for the anchored ZnPc on 2 ML NaCl/Ag(111). When compared with the results on 3 ML NaCl (Fig. 2), it is clear that the absolute energies of the Q(0,0) transition depend on the thickness of the decoupling layer, but the LSM does not.

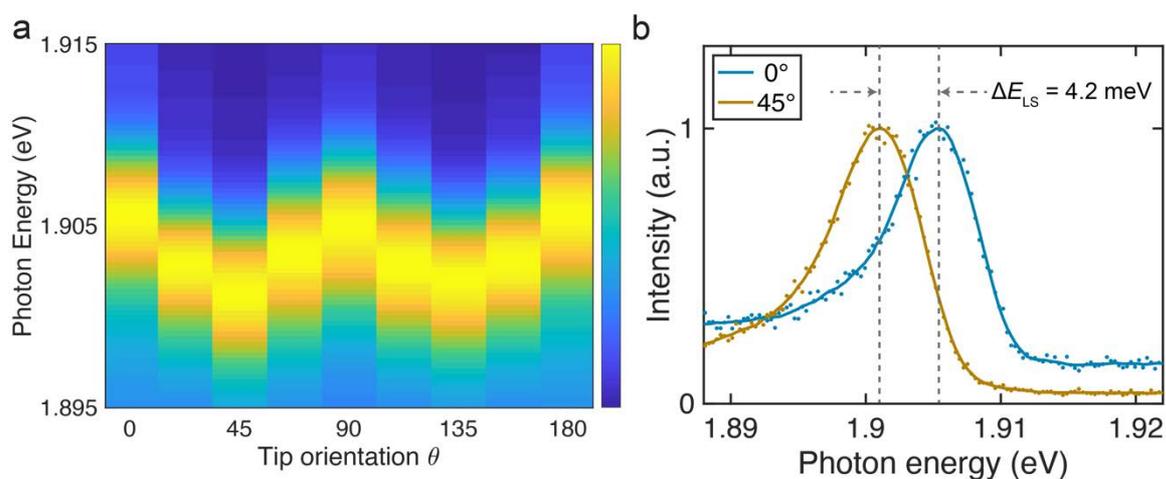

**Figure S6: Lamb shift modulation of anchored ZnPc on 2ML NaCl/Ag(111).** (a) Color plot of normalized STM-LE spectra showing periodic modulation of the Lamb shift. (b) Two typical Fano-peak STM-LE spectra taken on top of anchored ZnPc ($r$ = 915 pm) for $\theta$ = 0° and $\theta$ = 45° (cf. Fig. 1(c)), with $\Delta E_{LS}$= 4.2 ± 0.3 meV (taken in constant-current mode, $V_S$ = -2.5 V and $I_T$ = 100pA, $t$ = 180s). The solid lines are 20-point Savitzky-Golay filtered curves of the raw data.

**S5 Fitting STM-LE spectra by Fano profiles**

In general, a Fano profile reflects the coherent coupling between a discrete state and a continuum of states. The interplay between the inelastically tunneling electrons, the exciton of the molecule, and the plasmonic resonance has previously been suggested to fulfill these criteria[7,8]. Here, we fit our STM-LE spectra using the equation $F(E) = A \cdot f(E) + B$, where $A$ is a scaling factor, $B$ accounts for an offset background intensity, $E$ is the photon energy, and *f* is the Fano function[9]

$$f(q, \Gamma, E_0; E) = \frac{\left(q + \frac{E - E_0}{\Gamma}\right)^2}{1 + \left(\frac{E - E_0}{\Gamma}\right)^2}$$

with the Fano asymmetry parameter $q$, the resonant energy of the exciton $E_0$, and the linewidth (half width at half maximum) of the resonance $\Gamma$. The asymmetry parameter, *q*, roughly reflects the ratio of the different excitation channels responsible for forming the molecular exciton. Plasmon-induced exciton formation dominates for $|q| < 1$, leading to an asymmetric dip feature in the resulting STM-LE spectrum, as seen in the spectra taken near an isolated molecule (Fig. S7(a)). On the other hand, dominant exciton formation from electrons resonantly tunnel into/out of the molecule is reflected by $|q| > 1$, resulting in an asymmetric peak in the STM-LE spectrum. This is seen in the spectra taken on top of an anchored molecule (Fig. S7(b)). We did not observe a significant radius-dependent variation of $q$ when the tip was still located on top of the phenyl arm ($\theta = 45°$, Fig. S8) but a sudden jump to $|q| < 1$ for $r > 1.5$ nm. As all cases are nicely fitted by the Fano function, we conclude that the discrete exciton state coupled to the continuum of the plasmonic cavity fulfill the Fano criterion.

We note that a simple Fano profile is not sufficient to reproduce the spectrum taken on top of the isolated ZnPc due to the shoulder at higher energy. An additional peak was needed to fit the spectrum of the isolated ZnPc. Surprisingly, we were not able to adequately fit the shoulder with a second Fano profile, rather the best fit was made by a Gaussian peak (Fig. S7(c)). Although the underlying reason for the Gaussian lineshape is unclear, the fit enables the extraction of reliable values for the main transition peak (1.902 eV) and a good estimation for the position of the shoulder (1.911 eV). We note that the fit results of the main peak do not change significantly when fitting the shoulder with a Fano or a Lorentzian profile.

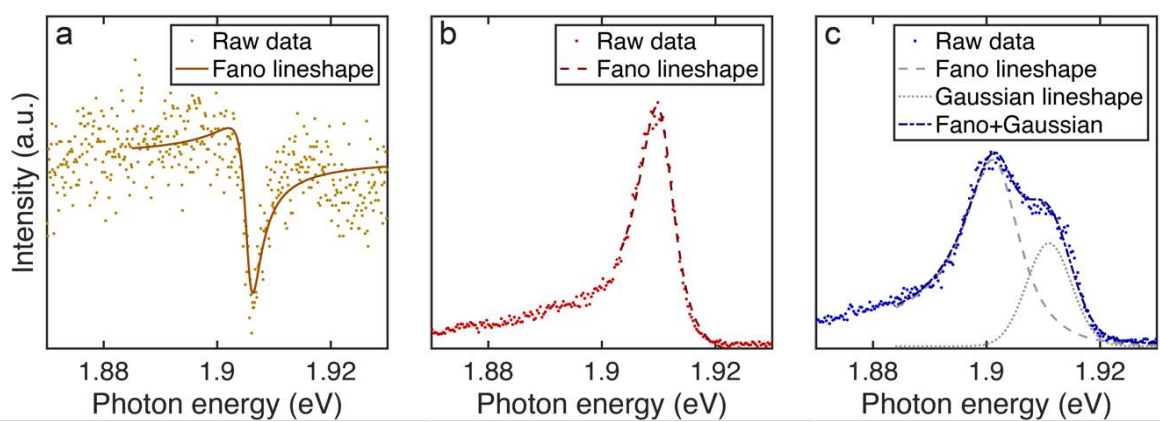

**Figure S7: Fano lineshape fitting.** STM-LE spectra taken at (a) close proximity to an isolated ZnPc (b) an anchored molecule, and (c) isolated ZnPc. Raw data are plotted as dots and the solid lines are the fitting results with asymmetry parameter $q$ = −0.4673 ± 0.0792, −4.884 ± 0.253, and −6.704 ± 1.782 respectively.

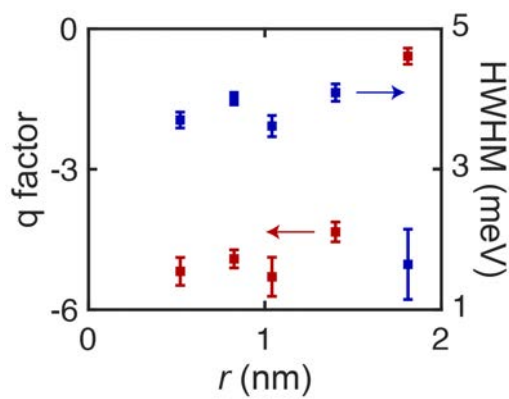

**Figure S8: Radius dependent Fano fitting parameters.** Radius dependence of the fitted Fano asymmetry parameter $q$ (red squares) and the fitted half width half maximum $\Gamma$ (blue squares). The azimuthal angle was fixed at 45° (along the phenyl arm). STM-LE spectra ($r$ = 520 pm) were taken in constant-height mode with feedback loop opened at the ZnPc center, $V_S$ = –2.5 V, $I$ = 150 pA, $t$ = 120 s. STM-LE spectra ($r$ = 825 pm, 1040 pm) were taken in constant-height mode with feedback loop opened at the ZnPc center, $V_S$ = –2.5 V, $I$ = 500 pA, $t$ = 120 s. STM-LE spectra ($r$ = 1400 pm) were taken in constant-current mode ($V_S$ = –2.5 V, $I$ = 200 pA, $t$ = 60 s). STM-LE spectra ($r$ =1810 pm) were taken in constant-current mode ($V_S$ = -2.5 V, $I$ = 200 pA, $t$ = 60 s).